\documentclass[a4paper,11pt]{article}
\usepackage{pos}

\title{Sensitivity to keV-MeV dark matter from cosmic-ray scattering with current and the upcoming ground-based arrays CTA and SWGO}
 \ShortTitle{Sensitivity to kev-MeV dark matter from cosmic-ray scattering}

\author*[a,b]{Igor Reis}
\author[b]{Emmanuel Moulin}
\author[a]{Aion Viana}

\affiliation[a]{Universidade de São Paulo, Instituto de Física de São Carlos,
  Av. Trabalhador São Carlense 400, São Carlos, Brazil}

\affiliation[b]{IRFU, CEA, Université Paris-Saclay, 
F-91191 Gif-sur-Yvette, France}
\onbehalf{for the CTA Consortium and SWGO Collaboration}

\emailAdd{igorreis@ifsc.usp.br}
\emailAdd{aion.viana@ifsc.usp.br}
\emailAdd{emmanuel.moulin@cea.fr}

\abstract{A wealth of astrophysical and cosmological observational evidence shows that the matter content of the universe is made of about 85$\%$ of non-baryonic dark matter. Huge experimental efforts have been deployed to look for the direct detection of dark matter via their scattering on target nucleons, their production in colliders, and their indirect detection via their annihilation products. Inelastic scattering of high-energy cosmic rays off dark matter particles populating the Milky Way halo would produce secondary gamma rays in the final state from the decay of the neutral pions produced in such interactions, providing a new avenue to probe dark matter properties. We compute here the sensitivity for H.E.S.S.-like observatory, a current-generation ground-based Cherenkov telescopes, to the expected gamma-ray flux from collisions of Galactic cosmic rays and dark matter in the center of the Milky Way. We also derive sensitivity prospects for the upcoming Cherenkov Telescope Array (CTA) and Southern Wide-field Gamma-ray Observatory (SWGO). The expected sensitivity allows us to probe a poorly-constrained range of dark matter masses so far, ranging from keV to sub-GeV, and provide complementary constraints on the dark matter-proton scattering cross section traditionally probed by deep underground direct dark matter experiments.}

\ConferenceLogo{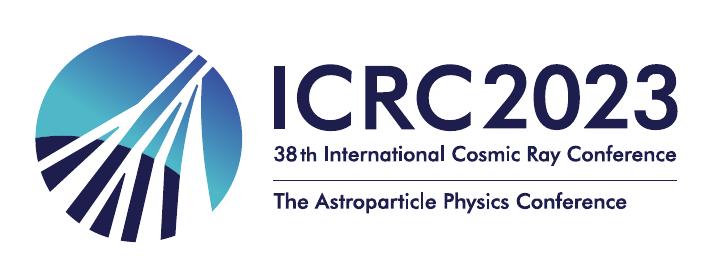}

\FullConference{%
38th International Cosmic Ray Conference (ICRC2023)\\
  26 July - 3 August, 2023\\
  Nagoya, Japan}

\begin{document}
\maketitle

\section{Introduction}
Very-high-energy (VHE, $\gtrsim$100 GeV) gamma-rays are crucial messengers for understanding non-thermal phenomena taking place in the Universe. The Galactic Center (GC) region stands out as one of the most studied regions of the sky over a large range of 
wavelengths. 
Many
gamma-ray observatories have focused on this rich and complex region with arrays of Imaging Atmospheric Cherenkov Telescopes (IACT) such as H.E.S.S. (see, for instance, Ref.~\cite{HESS:2022ygk} and references therein), as well as satellite experiments such as Fermi-LAT~\cite{Fermi-LAT:2017opo}, where
numerous gamma-ray emission have been detected. The Galactic Center region also harbors a high dark matter (DM) density~\cite{Hooper:2010mq} which makes it an ideal location
where great efforts have been deployed to search for dark matter signal signatures. 
The quest for dark matter in the form of Weakly Interacting Massive Particles (WIMP)~\cite{Bertone:2010zza} motivates worldwide experimental effort to probe their production at particle colliders~\cite{Kahlhoefer:2017dnp}, their scattering off nuclei in underground detectors~\cite{Schumann:2019eaa}, and 
their annihilation and decay via indirect detection~\cite{Strigari:2018utn}.

Observations of the Galactic Center region by H.E.S.S. have led to the strongest constraints to date for DM annihilation with masses in the TeV mass range~\cite{HESS:2022ygk}. 
Direct detection experiments have already heavily constrained the parameter space for ten-to-hundred GeV WIMP particles (see, for instance, Ref.~\cite{LZ:2022ufs}).  Despite huge efforts to probe 
DM interactions in GeV to hundred TeV mass range, DM remains elusive.
On the other hand, the sub-GeV mass range is comparatively poorly probed. Cosmological observations require the DM-proton scattering cross section $\sigma_{\chi p} \lesssim$ 10$^{-27}$ cm$^2$ for sub-GeV masses~\cite{Ali-Haimoud:2015pwa,Gluscevic:2017ywp,Xu:2018efh,Slatyer:2018aqg}.
This prompts scientists to consider other types of techniques to detect dark matter.
Here we focus on deriving sensitivity to sub-GeV DM via 
the gamma-ray signatures expected from the scattering of high-energy 
cosmic-rays (CR) off DM particles populating the DM halo of the Milky Way.

\section{Gamma rays from cosmic-ray boosted dark matter
}

\begin{figure}[htb]
    \centering
    \includegraphics[scale=0.15]{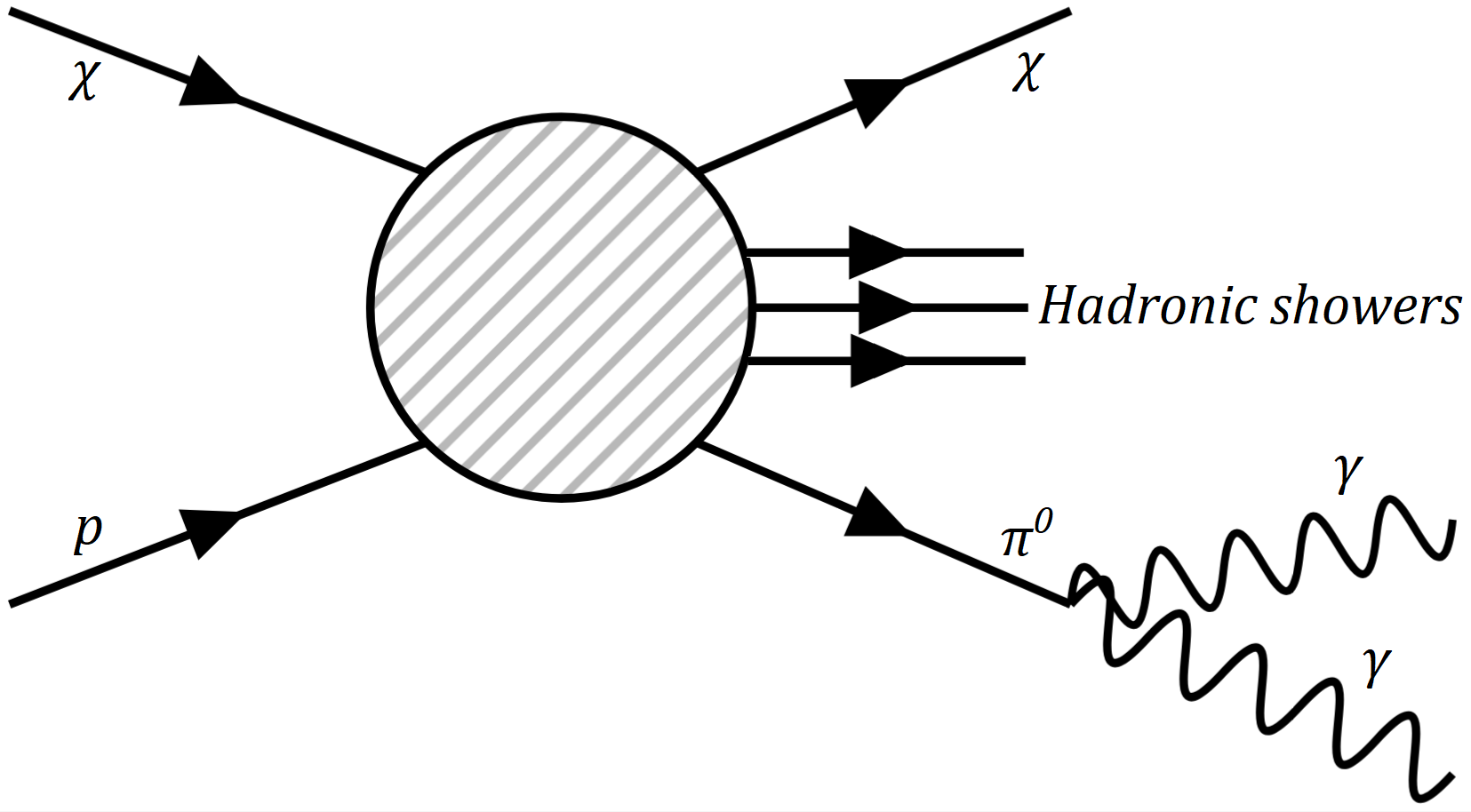}
    \caption{Schematics of the cosmic-ray proton scattering off a DM particle $\chi$ with production of gamma-rays from $\pi^0$ decay in the final state.}
    \label{fig:diagram}
\end{figure}

Dark matter particles $\chi$ in Galactic halo are inevitably upscattered by cosmic rays~\cite{Guo:2020oum, Cyburt:2002uw}. The subsequent scattering of the CR-boosted DM particles can probe elastic DM-nucleon cross section for light dark matter. Providing the center-of-mass energy of the CR-DM collision reaches GeV energies, deep inelastic scattering (DIS) will produce  gamma-rays and neutrinos in the final state from hadronization and decay of the standard
model particles produced in the interaction process. The inelastic process is expected to occur like in the $pp, p\nu, p\gamma$ collisions, independently of the nature of the DM particle.
Figure~\ref{fig:diagram} shows a scketch of the gamma-ray production 
from DIS of high-energy CR proton off the DM particle $\chi$ via  $p + \chi \rightarrow X + \chi \rightarrow \chi + {\rm hadronic\, showers} + \gamma$, where the final-state gamma rays originate from the $\pi_0$ decay.


The energy-differential gamma-ray flux expected from the scattering of protons of mass $m_{\rm p}$ off a DM particle of mass  $m_{\rm \chi}$ in a solid angle $\Delta\Omega$ can be expressed as 
\begin{equation}
    \frac{d\Phi_{\gamma}(E_{\gamma})}{dE_{\gamma}} = \frac{2}{m_{\chi}} \int_{E_{\rm min}}^{\infty}\frac{1}{f_{\rm \pi}}\sigma_{\rm \chi p}^{\rm inel}\left(\frac{E_{\rm p}}{f_{\rm \pi}}\right)\frac{d\phi_{\rm p}}{dE_{\rm p}}\left(\frac{E_{\rm p}}{f_{\rm \pi}}\right)dE_{\rm p}\times \int d\Omega\int_{l.o.s.}\rho_{\rm DM} [r(s)]ds \, ,
    \label{eq:flux}
\end{equation}
where 
\begin{equation}
    E_{\rm min} = (m_{\rm p} + m_{\rm \pi^{0}})\left[1 + \frac{m_{\rm \pi^{0}}(2m_{\rm p} + m_{\rm \pi^{0}})}{2m_{\rm \chi}(m_{\rm p} + m_{\rm \pi^{0}})}\right] \, .
    \label{eq:emin}
\end{equation}
In this equation we can clearly see the dependence on the DM halo density profile $\rho_{\rm DM}$. The integral over $\Delta\Omega$ and the line of sight (l.o.s) $s$ is usually referred as to the D-factor in the context of decaying DM searches.
$d\phi_{p}/dE_{p}$ corresponds to the CR flux in the regions where the scattering occurs and $\sigma_{\chi p}^{\rm inel}$ is the inelastic cross section.  
The energy transfer fraction between the incoming cosmic-rays and pions after the upscattering is given by $f_{\pi}$. The pion production process is an energy threshold interaction that requires a 
minimum CR energy $E_{\rm min}$ given in Eq.~(\ref{eq:emin}). 
 
The DM density distribution is parametrized by
the Einasto density profile

\begin{equation}
    \rho(r) = \rho_{s} \exp \left\{-\frac{2}{\alpha}\left[\left(\frac{r}{r_{s}}\right)^{\alpha} - 1\right]  \right\},
    \label{eq:einasto}
\end{equation}

\noindent where $\rho_{s}$ = 0.081 GeVcm$^{-3}$ and $r_{s}$ = 20 kpc are the scale density and radius, respectively, $\alpha = 0.17$ is a steepness index and $r$ is the radial coordinate from the center of the halo expressed $r = \big(s^2 +r_{\odot}^2-2\,r_{\odot}\,s\, \cos\theta \big)^{1/2}$, with $r_\odot$ is the distance of the observer to the GC taken to be
$r_\odot$ = 8.5 kpc, and $\theta$ corresponds to the angle between the direction of observation and the GC.

The energy-differential CR flux is assumed to follow 
a power law given by $d\phi_{\rm p}(E_{\rm p})/dE_{\rm p} = \Phi_0 (E_{\rm p}/1 {\rm TeV})\times E^{\rm -\Gamma}$. 
In the GC region, the VHE CR flux should be higher and harder than the local one since the CR density increases there compared to the CR see~\cite{HESS:2016pst,Gaggero:2017jts}. 
We conservatively set the flux normalization to the Solar-neighborhood value $\Phi_0$ = 10$^{-8}$ TeV$^{-1}$ cm$^{-2}$s$^{-1}$sr$^{-1}$, since it is known and has been measured 
and set the spectral index $\Gamma = 2.3$ as it has been measured by H.E.S.S.~\cite{HESS:2016pst}.

Following the neutrino-nucleon scattering cross section measurements, 
only slight energy-dependence is observed in the GeV energy range~\cite{Formaggio:2012cpf}.
In what follows we will assume no energy dependence of the cross section with energy. 
Since we are not assuming any specific underlying DM particle model, 
the inelastic and elastic cross-sections can be linked
as $\sigma^{\rm inel} = \alpha_2\sigma^{\rm el}$, where 
 both elastic and inelastic collisions are excitations of the internal degrees of freedom of the proton. Here relate both by an isospin factor $\alpha_{2} = 2/3$ \cite{Cyburt:2002uw}.

Given the $E_{\rm min}$ dependence on $m_{\rm \chi}$, we can derive a minimum dark matter mass detectable in a given experiment. For lower masses, we need higher minimum energy to produce a gamma-ray signal, and this implies that the detector would also need to have a good sensitivity for this high energy signal. So even though we are not imposing any constraints in the dark matter mass through any kind of particle physics model, a lower mass detection limit is imposed by the high energy resolution of the detection experiment we would use.
Following the parallel with the $pp$ inelastic scattering, we need a energy transfer fraction parameter $f_{\pi}$, that dictates how much of the kinetic energy of the cosmic rays is transferred to the produced neutral pions. From neutrino-nucleon scattering measurements, we use $f_{\pi} = 10\%$. 

\section{Analysis framework and sensitivity computation}

\subsection{VHE gamma-ray observatories}

Currently, H.E.S.S. 
is 
monitoring the GC. H.E.S.S. is the one with the highest sensitivity in the TeV energy range.
The next generation of gamma-ray telescopes are the Cherenkov Telescope Array (CTA), and the Southern wide field-of-view gamma-ray Observatory (SWGO), both are expected to probe high energies, at the TeV scale. They will be described in more detail in the following:

\begin{itemize}
    
    \item H.E.S.S. is a ground-based  array of five imaging atmospheric Cherenkov telescopes (IACT), located in the Khomas Highland of Namibia~\cite{hessweb} 
     it has been monitoring the GC 
    since 2003. Able to probe an energy interval of 230~GeV to 100~TeV, it provides the most stringent limits on the annihilation cross section probing thermal TeV WIMPs with the Inner Galaxy Survey programme~\cite{HESS:2022ygk}. For this work, we will use 500 hours live time looking at the inner $3^\circ$ region of the inner Galactic halo. It is important to say that we exclude the central region of the galaxy with $|b|<0.3^\circ$, to avoid the strong background that comes from this central region.   
    
    \item CTA: currently under construction, it is also an IACT array, the CTA will be situated on both hemispheres, the Northern site in La Palma, Canary Islands, Spain and the Southern site in Paranal, Chile. It is expected to probe an energy interval from $20$~GeV to $300$~TeV~\cite{CTA:2020qlo}, and it will be able, for the first time, to probe the region of the thermal averaged cross-section for DM masses at TeV-scales. In this study, we will use 500 hours of observation time, and also consider a circular $5^\circ$ 
    region around the GC, excluding also the Galactic latitude $\pm 0.3^\circ$~\cite{Moulin:2019oyc}.  
    
    \item SWGO: in the research and development phase, it is planned to be located at the South America. SWGO is a water Cherenkov particle detector array, with a wide field of view, and high duty cycle. It will have an energy sensitivity of $500$~GeV to $2$~PeV \cite{Albert:2019afb}. It is expected to have an unprecedented sensitivity to multi-TeV DM masses. In this work, we will use ten years of observation time, looking at $10^\circ$ of the central region of the Galactic center, again excluding the central region $|b|<0.3^\circ$.   
\end{itemize}

\subsection{Statistical data analysis and sensitivity computation} 
In order to assess the sensitivity of the current and future VHE gamma-ray observatories, we construct mock datasets. For H.E.S.S.-like and CTA,
we define 
the Regions of Interest (ROIs) as annuli centered on the GC of 0.1$^\circ$ width inside a 3$^\circ$ circular region for H.E.S.S., a $5^{\circ}$ circular region for CTA 
and of width 0.2$^\circ$ inside the inner 10$^\circ$ circular region for SWGO. 

\begin{figure}[htb]
    \centering
    \includegraphics[scale=0.7]{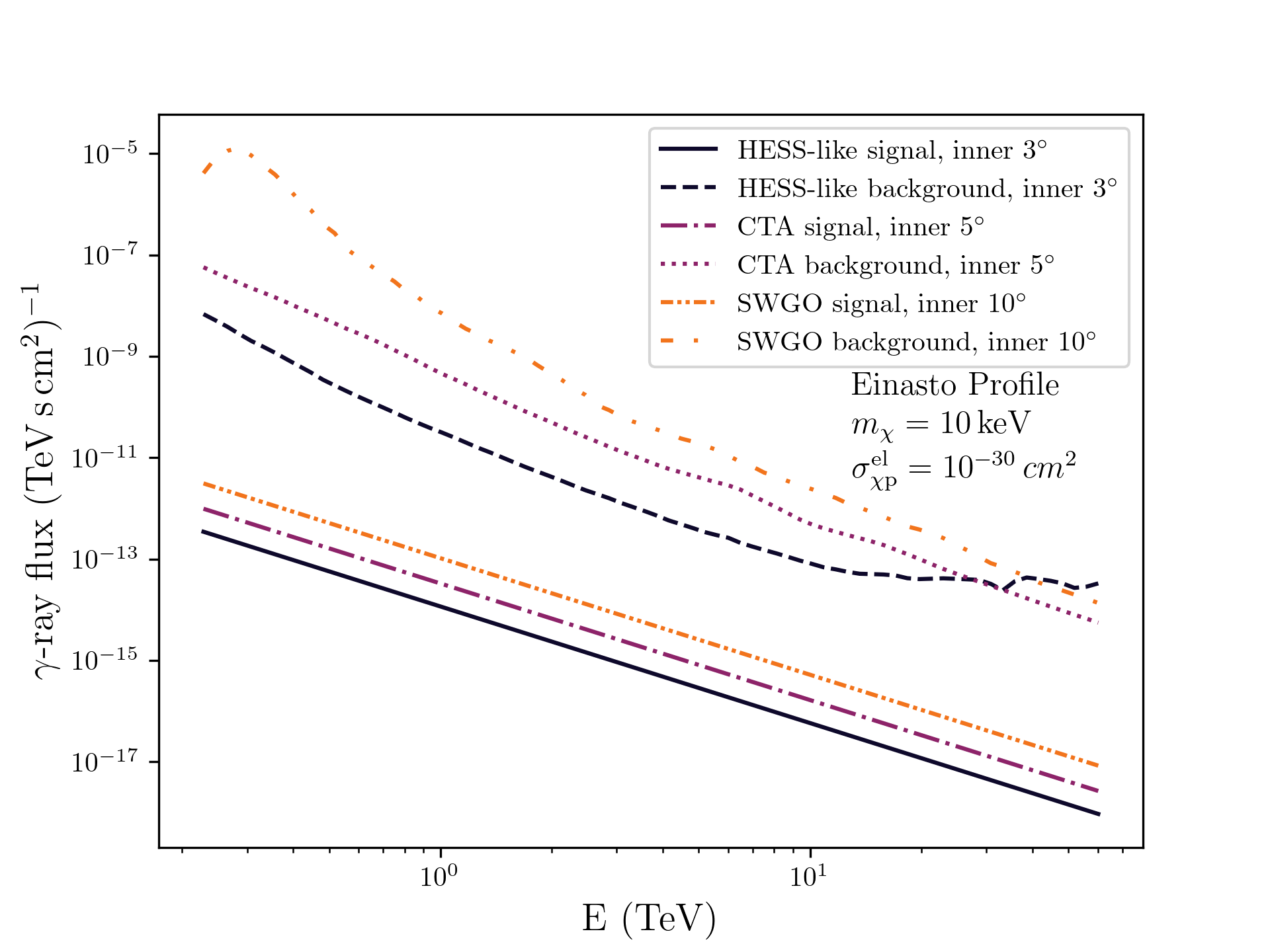}
    \caption{Gamma-ray fluxes expected for the DM-induced signal and residual background in the GC region for H.E.S.S.-like, CTA and SWGO observations. We
    show the expected signal spectra for a DM halo following the Einasto density profile populated by DM particle of mass $m_{\rm \chi}$ =  10 keV. The $p\chi$ elastic cross section is taken to be $\sigma^{\rm el}_{\rm \chi p}$ = 10$^{-30}$ cm$^{2}$. The CR spectrum follows a power law with spectral index of $\Gamma$ = 2.3. For each observatory, the background spectrum corresponds to the residual background. See text for more details.}
    \label{fig:flux}
\end{figure}

The analysis is performed in an ON/OFF framework (see, for instance, Ref.~\cite{HESS:2022ygk}). The signal events are searched in the ON region consisting of the ROIs, 
and the background events taken from OFF regions are obtained from 
Monte Carlo simulations of the expected background.
Then, the data analysis itself is performed as using a 2D-binned Poisson maximum log-likelihood ratio test statistics. 
Our data are binned into spatial and energy bins. The spatial bins are taken as the different ROIs, and for the energy binning, we divided the whole energy range of the experiments into 92 logarithmically-spaced bins. The likelihood in the $i$ energy and $j$ spatial bins writes as:
\begin{equation}
    \mathcal{L} = \frac{(N_{\rm S,ij} + N_{\rm B,ij})^{N_{\rm ON,ij}}}{N_{\rm ON,ij}!}e^{-(N_{\rm S,ij} + N_{\rm B,ij})} \times \frac{(N'_{\rm S,ij}+N_{\rm B,ij})^{N_{\rm OFF,ij}}}{N_{\rm OFF,ij}!}e^{-(N'_{\rm S,ij}+N_{\rm B,ij})}\, .
    \label{eq:liklh}
\end{equation}
$N_{\rm ON,ij}$ and $N_{\rm OFF,ij}$ are the numbers of events in the ON and OFF regions, respectively. $N_{\rm S,ij} + N_{\rm B,ij}$ represents the expected number of events in the $i,j$ energy and spatial bins in the ON region, with $N_{\rm S,ij}$ being the expected signal.
$N_{\rm B,ij}$ is extracted from the OFF region. 
Similarly, $N'_{\rm S,ij} + N_{\rm B,ij}$ corresponds to the expected number of events in the $i,j$ energy and spatial bins in the OFF region.
The number of expected signal events is obtained by folding the expected gamma-ray flux given in Eq.~(\ref{eq:flux}) with the energy-dependent acceptance of the instrument. The latter are extracted for H.E.S.S.-like~\cite{HESS:2022ygk}, CTA~\cite{CTAOirf} and SWGO~\cite{swgoirf} observatories, respectively.
We assumed that the OFF regions are taken far away from the ON regions, 
such that we can neglect the amount of expected signal there, \textit{i.e.}, $N'_{\rm S,ij}$ = 0.

In order to constrain $\sigma^{el}_{\chi p}$ as a function of the DM particle mass, we can obtain a joint likelihood function $\mathcal{L}$ multiplying the Poisson likelihoods of each spatial \textit{i} and energy \textit{j} bins $\mathcal{L}_{ij}$ such as $\mathcal{L}\left(m_{\rm \chi}, \sigma^{el}_{\chi p}\right) = \prod_{ij}\mathcal{L}_{ij}$.
In order to compute the sensitivity on the elastic cross section, we perform a likelihood ratio test statistics $TS = -2\ln{\left( \mathcal{L}_{0}/\mathcal{L}_{max}\right)}$, maximizing the joint likelihood, where $\mathcal{L}_{0}$ is the null hypothesis, where we have no DM signal. This TS follows a $\chi^{2}$ distribution in the high statistic regime, and therefore allows us to compute the mean expected upper limit  on the cross section at 95\% confidence level (C.L.) by imposing 
TS = 2.71. 
For the computation of the expected limits, we follow the Asimov procedure~\cite{Cowan:2010js}. Instead of generating many realizations of the expected background, the mean expected background is taken as the data to compute the mean of the expected sensitivity (see, for instance, Ref.~\cite{Montanari:2022buj} for more details).

\section{Results and discussion}
\begin{figure}[htb]
    \centering
    \includegraphics[scale = 0.7]{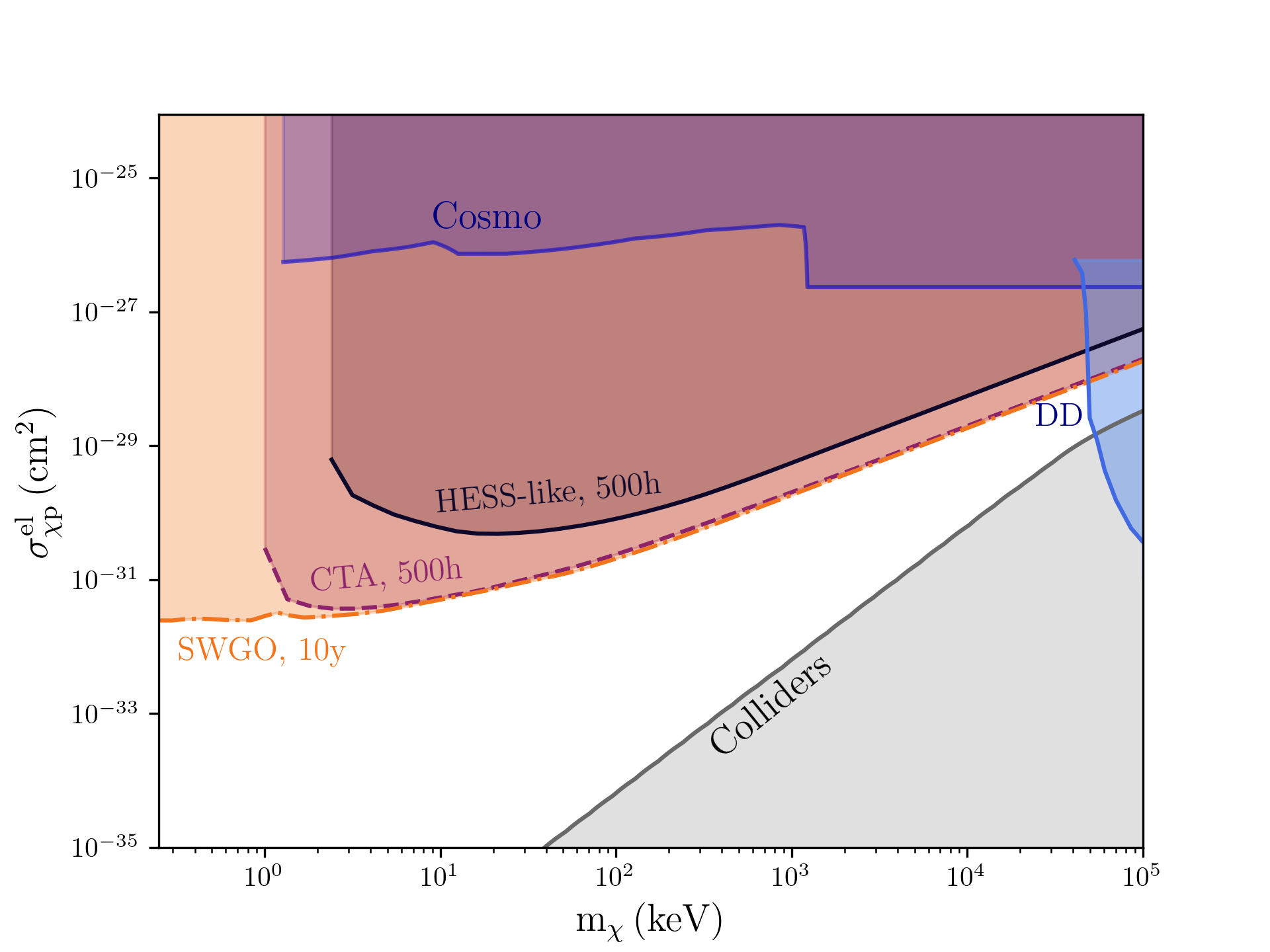}
    \caption{Comparison between the sensitivity reach derived in this work with current direct detection (DD) experiment constraints, colliders and cosmological searches for the elastic cross section between DM and protons. Adapted from Ref.~\cite{Cappiello:2018hsu}.}
    \label{fig:limcomp}
\end{figure}

We compute expected upper limits for the elastic cross section $\sigma^{\rm el}_{\chi p}$ between energetic CRs and DM particles at 95$\%$ confidence level to derive the sensitivity of the current and planned VHE gamma-ray observatories. Assuming the DM distribution in the Milky Way to follow an Einasto density profile, the sensitivity expected for each experiment are shown in Figure
~\ref{fig:limcomp}, for DM masses ranging from 
a fraction of keV up to a 100 MeV. 
The effect of energy threshold of the reaction process combined with the energy-dependent behaviour of the acceptance is observed for H.E.S.S.-like and CTA observatories due to the drop in acceptances for the lowest accessible energies. However, in the case of SWGO the threshold effect is hidden due to large fluctuations in the expected residual background estimate~\cite{Viana:2019ucn} used in this work.
The strongest sensitivity reaches $3.6\times 10^{-32}$cm$^{2}$ and $4.8\times 10^{-31}$cm$^{2}$ for a DM mass of  2 keV 
and 21 keV for CTA and H.E.S.S.-like, respectively. Due to the SWGO expected flux sensitivity up to 2 PeV, lower masses can be probed, reaching $2.4\times 10^{-32}$cm$^{2}$ for a DM mass of $0.2\,\rm keV$.

Figure~\ref{fig:limcomp} also presents our results in the context of current constraints
in the keV-to-MeV mass range. A significant fraction
of an unexplored region of the parameter space can be probed by VHE gamma-ray observatories where we reach here sensitivity of about 10$^{-30}$ cm$^{2}$ for keV-MeV DM masses. 
If we compare the sensitivity of VHE observatories  
with the limits obtained with direct detection experiments \cite{Billard:2021uyg, Cappiello:2018hsu}, 
the indirect method 
making use of CR scattering off DM particles shows itself as a way to probe a currently-unexplored region of the elastic scattering space parameter. 


\section{Acknowledgments}

This work was conducted in the context of the CTA Consortium\footnote{\url{https://www.cta-observatory.org/consortium_acknowledgments/}} and the SWGO Collaboration. It has made use of the CTA instrument response functions provided by the CTA Consortium and Observatory, see https://www.ctao-observatory.org/science/cta-performance/version prod5 v0.1~\cite{CTAOirf}, and the SWGO instrument response functions provided by the SWGO Consortium (see~\cite{swgoirf}). 
This work is supported by the "ADI 2021" project funded by the IDEX Paris-Saclay, ANR-11-IDEX-0003-02, and by FAPESP, process numbers 2019/14893-3, 2021/02027-0 and 2021/01089-1. AV is supported by CNPq grant {\rm 314955/2021-6}.

%
%
%

\end{document}